\documentclass[prd,twocolumn,groupedaddress,showpacs,nofootinbib]{revtex4}
\usepackage{graphicx}
\usepackage{dcolumn}
\usepackage{amssymb}
\usepackage{mathrsfs}
\usepackage{amsmath}
\usepackage{epsfig}
\usepackage[dvips]{color}
\usepackage{hhline}

\begin{document}

\title{Origin of CP violation for leptogenesis in seesaw}

\author{Pei-Hong Gu}

\email{peihong.gu@sjtu.edu.cn}

\affiliation{Department of Physics and Astronomy, Shanghai Jiao Tong University, 800 Dongchuan Road, Shanghai 200240, China}

\begin{abstract}

We reveal the origin of the CP violation required by the leptogenesis in variously popular seesaw models. Especially we clarify that in a pure type-I/III seesaw with two fermion singlets/triplets, a combined type-I+III seesaw with one fermion singlet and one fermion triplet, or a combined type-I/III+II seesaw with one fermion singlet/triplet and one Higgs triplet, the CP violation for the leptogenesis can exactly come from the imaginary part of the neutrino mass matrix in a special basis where the Yukawa couplings involving one fermion singlet/triplet are allowed to get rid of any CP phases. We also generalize our findings as a very good approximation when these seesaw scenarios are extended by more fermion singlets/triplets and Higgs triplets while the leptogenesis is realized by the decays of the lightest fermion singlet/triplet.

\end{abstract}


\maketitle

\underline{Introduction:} The atmospheric, solar, accelerator and reactor neutrino experiments have established the phenomena of neutrino oscillations \cite{patrignani2016}. This requires a mixing among three flavors of massive neutrinos and hence a necessity for new physics beyond the standard model (SM). Meanwhile, the cosmological observations have indicated that the neutrino masses should be in a sub-eV range \cite{patrignani2016}. In order to naturally explain the smallness of the neutrino masses, we can resort to the famous seesaw mechanism \cite{minkowski1977,yanagida1979,grs1979,ms1980}. The essential feature of the seesaw mechanism is that the neutrino masses can be highly suppressed by a small ratio of the electroweak scale over a newly high scale. Currently, the most popular seesaw models include the so-called type-I \cite{minkowski1977,yanagida1979,grs1979,ms1980}, type-II \cite{mw1980,sv1980,cl1980,lsw1981,ms1981} and type-III \cite{flhj1989} seesaw. The type-I/III seesaw is realized by introducing some fermion singlets/triplets with a heavy Majorana mass term as well as the Yukawa couplings to the SM lepton and Higgs doublets. As for the type-II seesaw, it contains some heavy Higgs triplets with the Yukawa couplings to the SM lepton doublets as well as the cubic couplings to the SM Higgs doublet.

Remarkably, these seesaw models can also accommodate a leptogenesis \cite{fy1986,lpy1986,luty1992,mz1992,fps1995,crv1996,pilaftsis1997,ms1998,hs2004,ak2004} mechanism to solve the puzzle of the cosmic matter-antimatter asymmetry, which is equivalent to a baryon asymmetry. In this seesaw and leptogenesis scenario, the neutrino mass and the baryon asymmetry can be simultaneously induced by certain interactions involving the newly heavy particles. However, such seesaw models contain many free parameters. This leads to a conventional wisdom that the corresponding leptogenesis cannot give a distinct relation between the baryon asymmetry and the neutrino mass matrix unless we do some assumptions on the texture of the relevant masses and couplings. For example, ones can expect a successful leptogenesis in the canonical type-I seesaw model even if the neutrino mass matrix does not contain any CP phases \cite{di2001}.

In this work we shall reveal that in a pure type-I/III seesaw with two fermion singlets/triplets, a combined type-I+III seesaw with one fermion singlet and one fermion triplet, or a combined type-I/III+II seesaw with one fermion singlet/triplet and one Higgs triplet, the CP violation required by the leptogenesis exactly originates from the imaginary part of the neutrino mass matrix. This is because for a special basis, the Yukawa couplings involving one of the two fermion singlets/triplets or the unique Higgs triplet are always allowed to absorb all of the physical CP phases in the lepton sector. We shall also clarify that in the seesaw models with more fermion singlets/triplets and Higgs triplets, the imaginary part of the neutrino matrix approximately is the source of the CP violation for the leptogenesis by the decays of the lightest fermion singlet/triplet.

\underline{The type-I/II/III seesaw models:} For simplicity, we do not write down the full SM Lagrangian. Instead, we only show the part of the lepton sector, i.e.
\begin{eqnarray}
\label{sm}
\mathcal{L}_{\textrm{SM}}^{}&\supset& \sum_{\alpha}^{}\left[i \bar{l}_{L\alpha}^{} D \!\!\!\! /\, l_{L\alpha}^{}+ i \bar{e}_{R\alpha}^{} D \!\!\!\! /\, e_{R\alpha}^{}
- y_{\alpha}\left(\bar{l}_{L\alpha}^{}\tilde{\phi}e_{R\alpha}^{}\right.\right.\nonumber\\
[2mm]
&&\left.\left.+\textrm{H.c.}\right)\right]\,,
\end{eqnarray}
where $\phi$, $l_{L\alpha}^{}$ and $e_{R\alpha}^{}~(\alpha=e,\mu,\tau)$ respectively stand for the Higgs scalar, the left-handed leptons and the right-handed leptons, i.e.
\begin{eqnarray}
&&\begin{array}{l} \phi(1,2,-\frac{1}{2})\end{array}=\left[\begin{array}{l}\phi^{0}_{}\\
[1mm]
\phi^{-}_{}\end{array}\right]\,,
\quad 
\begin{array}{l} l^{}_{L\alpha}(1,2,-\frac{1}{2})\end{array}=\left[\begin{array}{l}\nu^{}_{L\alpha}\\
[1mm]
e^{}_{L\alpha}\end{array}\right]\,,\nonumber\\
[2mm]
&& \begin{array}{l} e^{}_{R\alpha}(1,1,-1)\,. \end{array}
\end{eqnarray}
Here and thereafter the brackets following the fields describe the transformations under the SM $SU(3)_c^{}\times SU(2)_L^{}\times U(1)^{}_{Y}$ gauge groups. Note we have taken the Yukawa couplings in Eq. (\ref{sm}) to be real and diagonal without loss of generality and for convenience.

We then review the most general type-I/II/III seesaw \cite{minkowski1977,yanagida1979,grs1979,ms1980,mw1980,sv1980,cl1980,lsw1981,ms1981,flhj1989},
\begin{subequations}
\begin{eqnarray}
\label{type-i}
\mathcal{L}_{\textrm{I}}^{}&=&\sum_{i}^{}\left [i \bar{N}_{Ri}^{} \partial \!\!\! / N_{Ri}^{}-\frac{1}{2}M_{N_i}^{} \left(\bar{N}_{Ri}^{} N_{Ri}^{c} +\textrm{H.c.}\right) \right.\nonumber\\
[1mm]
&&\left.-\sum_{\alpha}^{}\left(g_{\alpha i}^{}\bar{l}_{L\alpha i}^{}\phi N_{Ri}^{}+\textrm{H.c.}\right)\right]\,,\\
 [2mm]
 \label{type-ii}
\mathcal{L}_{\textrm{II}}^{}&=&\sum_{i}^{}\left\{\textrm{Tr}\left[\left(D_\mu^{}\Delta_i^{} \right)^\dagger_{}\left(D^\mu_{} \Delta_i^{} \right)\right]-M_{\Delta_i}^2 \textrm{Tr}\left(\Delta^\dagger_{i}\Delta_i^{} \right)\right.\nonumber\\
[1mm]
&&
- \frac{1}{2}\mu_i^{} \left(\phi^\dagger_{} \Delta_i^{} i \tau_2^{} \phi^\ast_{}+\textrm{H.c.}\right)-(\textrm{quartic~terms})\nonumber\\
[1mm]
&&
\left.-\sum_{\alpha\beta}^{}\left(\frac{1}{2}f_{\alpha\beta i}^{}\bar{l}_{L\alpha}^{}\Delta_i^{} i\tau_2^{} l_{L\beta}^c  +\textrm{H.c.}\right) \right\}\,,\\
[2mm]
\label{type-iii}
\mathcal{L}_{\textrm{III}}^{}&=&\sum_{i}^{}\left\{i\textrm{Tr}\left(\bar{T}_{Li}^{} D \!\!\!\!/ \,T_{Li}^{}\right)\right.\nonumber\\
[1mm]
&&-\frac{1}{2}M_{T_i}^{}\left[\textrm{Tr}\left( \bar{T}_{Li}^c i\tau_2^{} T_{Li}^{} i\tau_2^{}\right)+\textrm{H.c.}\right]\nonumber\\
[1mm]
&&\left.-\sum_{\alpha}^{}\left(\sqrt{2}\,h_{\alpha i}^{}\bar{l}_{L\alpha}^{}i\tau_2^{} T_{Li}^c i\tau_2^{} \phi +\textrm{H.c.}\right)\right\}\,,
 \end{eqnarray}
 \end{subequations}
where $N_{Ri}^{}$, $\Delta_i^{}$ and $T_{Li}^{}~(i=1,...,n\geq 1)$ respectively denote the fermion singlet(s), the Higgs triplet(s) and the fermion triplet(s), i.e. 
\begin{subequations}
 \begin{eqnarray}
N_{Ri}^{}(1,1,0)\,,&&\\
[2mm]
 \Delta_i^{} (1,3,-1)&=&\left[\begin{array}{ll}\delta^{-}_{i}/ \sqrt{2} & ~~\delta^{0}_{i}\\
[1mm]
\delta^{--}_{i}& -\delta^{-}_{i}/\sqrt{2}\end{array}\right]\,,\\
[2mm]
 T^{}_{Li}(1,3,0)&=&\left[\begin{array}{ll}T^{0}_{Li}/ \sqrt{2} & ~~T^{+}_{Li}\\
[1mm]
T^{-}_{Li}& -T^{0}_{Li}/\sqrt{2}\end{array}\right]\,.
\end{eqnarray} 
\end{subequations}
In the above Lagrangians, the CP phases only exist in the Yukawa couplings involving the fermion singlet(s)/triplet(s) and the Higgs triplet(s). This can be always achieved by a proper phase rotation.

It is easy to see that in a type-I/III, type-I+III or type-I/III+II seesaw extension of the SM, the Yukawa couplings involving one fermion singlet/triplet can be further chosen to be real.
In other words, all of the CP phases in the lepton sector can be included in the Yukawa couplings of the other fermion singlet(s)/triplet(s) to the lepton and Higgs doublets, and/or the Yukawa couplings of the Higgs triplet(s) to the lepton doublets. For the following demonstration, we conveniently assign
\begin{eqnarray}
\label{real}
g_{\alpha 1}^{}\equiv g^{\ast}_{\alpha 1} \quad \textrm{or}\quad h_{\alpha 1}^{}\equiv h^{\ast}_{\alpha 1}\,.
\end{eqnarray}
Actually, the above assignment can be understood by the phase rotation as below,
 \begin{eqnarray}
 \label{real2}
&&X\, g \rightarrow g  \quad \textrm{or}\quad X\, h \rightarrow h  \nonumber\\
[2mm]
&& \textrm{with}\quad X=\textrm{diag}\{e^{i\beta_1^{}}_{},~e^{i\beta_2^{}}_{},~e^{i\beta_3^{}}_{}\}\,.
\end{eqnarray} 
 
 \underline{The neutrino mass matrix and its CP phases:} When the Higgs scalar $\phi$ develops its VEV $\langle\phi\rangle=\langle\phi^0_{}\rangle=v\simeq 174\,\textrm{GeV}$ to spontaneously break the electroweak symmetry, the left-handed neutrinos $\nu_L^{}$ can acquire a tiny Majorana mass term by integrating out the heavy fermion singlet(s)/triplet(s) and/or Higgs triplet(s), i.e.
 \begin{eqnarray} \label{numass}
 \mathcal{L}&\supset& -\frac{1}{2} \bar{\nu}_{L}^{} m_\nu^{} \nu_{L}^c + \textrm{H.c.} ~~ \textrm{with}\nonumber\\
 [2mm]
 &&m_\nu^{}=m_\nu^{\textrm{I}}+m_\nu^{\textrm{II}}+m_\nu^{\textrm{III}} =X\,U\,\hat{m}\,U^T_{}\,X_{}^{T}\,.
  \end{eqnarray}
Here the mass matrices $m_\nu^{\textrm{I/II/III}} $ are the type-I/II/III seesaw \cite{minkowski1977,yanagida1979,grs1979,ms1980,mw1980,sv1980,cl1980,lsw1981,ms1981,flhj1989},
\begin{subequations}
\label{seesaw}
\begin{eqnarray}
\label{seesaw-i}
 \left(m_\nu^{\textrm{I}}\right)_{\alpha\beta}^{}&=&- \sum_{i}^{}g_{\alpha i}^{}g_{\beta i}^{}\frac{v^2_{}}{M_{N_i}^{}}\,,\\
 [2mm]
 \label{seesaw-ii}
 \left(m_\nu^{\textrm{II}}\right)_{\alpha\beta}^{}&=&-\sum_{i}^{}f_{\alpha\beta i}^{}\frac{\mu_i^{} v_{}^2}{2M_{\Delta_i}^2} \,,\\
 [2mm]
 \label{seesaw-iii}
 \left(m_\nu^{\textrm{III}}\right)_{\alpha\beta}^{}&=&- \sum_{i}^{}h_{\alpha i}^{}h_{\beta i}^{}\frac{v^2_{}}{M_{T_i}^{}}\,,
\end{eqnarray}
\end{subequations}
the diagonal matrix $\hat{m}$ gives three neutrino mass eigenvalues,
\begin{eqnarray}\label{pmns}
\hat{m} =\textrm{diag}\left\{m_1^{}\,,~m_2^{}\,,~m_3^{}\right\}\,,
\end{eqnarray}
while the PMNS matrix $U$ determines the mixing among three neutrino flavors \cite{patrignani2016}, 
\begin{widetext}
\begin{eqnarray}\label{pmns}
U=\left[\begin{array}{ccccl}
c_{12}^{}c_{13}^{}&& s_{12}^{}c_{13}^{}&&  s_{13}^{}e^{-i\delta}_{}\\
[1mm] -s_{12}^{}c_{23}^{}-c_{12}^{}s_{23}^{}s_{13}^{}e^{i\delta}_{}
&&~~c_{12}^{}c_{23}^{}-s_{12}^{}s_{23}^{}s_{13}^{}e^{i\delta}_{}
&& s_{23}^{}c_{13}^{}\\
[1mm] ~~s_{12}^{}s_{23}^{}-c_{12}^{}c_{23}^{}s_{13}^{}e^{i\delta}_{}
&& -c_{12}^{}s_{23}^{}-s_{12}^{}c_{23}^{}s_{13}^{}e^{i\delta}_{}
&& c_{23}^{}c_{13}^{}
\end{array}\right]\times \textrm{diag}\left\{e^{i\alpha_1^{}/2}_{}\,,~e^{i\alpha_2^{}/2}_{}\,,~1\right\}\,.
\end{eqnarray}
\end{widetext}

Clearly the neutrino mass matrix $m_\nu^{}$ is allowed to contain three physical CP phases: two Majorana phases $\alpha_{1,2}^{}$ and one Dirac phase $\delta$. The physical CP phases $\alpha_{1,2}^{}$ and $\delta$ as well as the unphysical CP phases $\beta_{1,2,3}^{}$ can appear if and only if there are some complex Yukawa couplings in the seesaw formula (\ref{seesaw}). By inserting the assignment (\ref{real}) into the seesaw formula (\ref{seesaw}), we conclude in the pure type-I/III seesaw, the combined type-I+III seesaw or the combined type-I/III+II seesaw, one fermion singlet/triplet will never contribute to the imaginary part of the neutrino mass matrix, i.e.
 \begin{itemize}
\item in the type-I seesaw,
 \begin{eqnarray}
\textrm{Im}\left[\left(m_\nu^{}\right)_{\alpha\beta}^{}\right]&=&- \textrm{Im}\left(\sum_{i\neq 1 }^{}g_{\alpha i}^{}g_{\beta i}^{}\frac{v^2_{}}{M_{N_i}^{}}\right)\,,
\end{eqnarray}
 \item in the type-III seesaw,
 \begin{eqnarray}
 \textrm{Im}\left[\left(m_\nu^{}\right)_{\alpha\beta}^{}\right]&=&-  \textrm{Im}\left(\sum_{i\neq 1 }^{}h_{\alpha i}^{}h_{\beta i}^{}\frac{v^2_{}}{M_{T_i}^{}}\right)\,,
\end{eqnarray}
\item in the type-I+III seesaw,
\begin{eqnarray}
&&\textrm{Im}\left[\left(m_\nu^{}\right)_{\alpha\beta}^{}\right]\nonumber\\
[1mm]
&=&-  \textrm{Im}\left(\sum_{i\neq 1 }^{}g_{\alpha i}^{}g_{\beta i}^{}\frac{v^2_{}}{M_{N_i}^{}}+ \sum_{j }^{}h_{\alpha j}^{}h_{\beta j}^{}\frac{v^2_{}}{M_{T_j}^{}}\right)\,,\nonumber\\
[2mm]
&\textrm{or}&\textrm{Im}\left[\left(m_\nu^{}\right)_{\alpha\beta}^{}\right]\nonumber\\
[1mm]
&=&-  \textrm{Im}\left(\sum_{i}^{}g_{\alpha i}^{}g_{\beta i}^{}\frac{v^2_{}}{M_{N_i}^{}}+ \sum_{j\neq 1}^{}h_{\alpha j}^{}h_{\beta j}^{}\frac{v^2_{}}{M_{T_j}^{}}\right)\,,
\end{eqnarray}
\item in the type-I+II seesaw,
\begin{eqnarray}
&&\textrm{Im}\left[\left(m_\nu^{}\right)_{\alpha\beta}^{}\right]\nonumber\\
[1mm]&=&-  \textrm{Im}\left(\sum_{i\neq 1 }^{}g_{\alpha i}^{}g_{\beta i}^{}\frac{v^2_{}}{M_{N_i}^{}}+ \sum_{j }^{}f_{\alpha\beta j}^{}\frac{\mu_j^{}v^2_{}}{2M_{\Delta_j}^{2}}\right)\,,
\end{eqnarray}
\item in the type-III+II seesaw,
\begin{eqnarray}
&&\textrm{Im}\left[\left(m_\nu^{}\right)_{\alpha\beta}^{}\right]\nonumber\\
[1mm]
&=&-  \textrm{Im}\left(\sum_{i\neq 1 }^{}h_{\alpha i}^{} h_{\beta i}^{} \frac{v^2_{}}{M_{T_i}^{}}+\sum_{j }^{}f_{\alpha\beta j}^{}\frac{\mu_j^{} v^2_{}}{2M_{\Delta_j}^{2}}\right)\,.
\end{eqnarray}
\end {itemize}

\underline{The CP violation for leptogenesis:} Either the fermion singlet(s)/triplet(s) or the Higgs triplet(s) or both can decay to generate a lepton asymmetry as long as the CP is not conserved. This lepton asymmetry then can partially get converted to a baryon asymmetry through the sphaleron processes \cite{krs1985}. Specifically, the final baryon asymmetry can be described by \cite{kt1990}
\begin{eqnarray}
\eta_B^{}\!\!&=&\!\! c_{\textrm{sph}}^{}\frac{\sum_{i}^{}\kappa_{N_i}^{}\varepsilon_{N_i}^{}+\sum_{j}^{}\kappa_{\Delta_j}^{} \varepsilon_{\Delta_j}^{} r_{\Delta_j}^{}+\sum_{k}^{}\kappa_{T_k}^{} \varepsilon_{T_k}^{} r_{T_k}^{}}{g_\ast^{}}\,.\nonumber\\
[0mm]
&&\!\!
\end{eqnarray}
Here $c_{\textrm{sph}}^{}=-\frac{28}{79}$ \cite{ht1990} is the sphaleron lepton-to-baryon coefficient, $g_\ast^{}=106.75$ \cite{kt1990} is the relativistic degrees of freedom during the leptogenesis epoch, $\kappa_{N_i/\Delta_i/T_i}^{}\lesssim 1$ denote the washout factors and their exact numbers are solved by the related Boltzmann equations, $r_{\Delta_i/T_i}^{}=3$ appear for the triplets, while $\varepsilon_{N_i/\Delta_i/T_i}^{}$ are the CP asymmetries in the decays of the fermion singlet $N_{i}^{}=N_{Ri}^{}+(N_{Ri}^{})^c_{}=N_{i}^c$, the Higgs triplet pair $(\Delta_{i}^{},~\Delta_{i}^{\ast})$ and the fermion triplet $T_{i}^{}=(T_i^{-},~T_i^0,~T_i^{+})$ with $T^0_{i}=T^0_{Li}+(T^0_{Li})^c_{}=(T^0_{i})^c_{}$ and $T^{\pm}_{i}=T^\pm_{Li}+(T^\mp_{Li})^c_{}=(T^\mp_{i})^c_{}$. The CP asymmetries $\varepsilon_{N_i/\Delta_i/T_i}^{}$ well characterize the CP violation required by the leptogenesis and they are evaluated at one-loop level \cite{fy1986,lpy1986,luty1992,mz1992,fps1995,crv1996,pilaftsis1997,ms1998,hs2004,ak2004},
\begin{widetext}
\begin{eqnarray}
\label{cpa-i}
\varepsilon_{N_i}^{}&=&\frac{1}{8\pi}\frac{\sum_{\alpha \beta}^{} \textrm{Im}\left\{g_{\alpha i}^\ast g_{\beta i}^\ast\left[ \sum_{j\neq i}  g_{\alpha j}^{}g_{\beta j}^{}I_{N_i}^{N_j} +\sum_{k}^{} f_{\alpha\beta k}^{} \frac{\mu_k^{}}{2M_{N_i}^{}}  I_{N_i}^{\Delta_k} 
+\sum_{l}^{}h_{\alpha l} h_{\beta l}^{}  I_{N_i}^{T_l} \right]  \right\}}{\sum_{\alpha}^{}g_{\alpha i}^\ast g_{\alpha i}^{} }\,,\\
[2mm]
\label{cpa-ii}
\varepsilon_{\Delta_i}^{}&=&\frac{1}{8\pi}\frac{\sum_{\alpha \beta}^{} \textrm{Im}\left\{f_{\alpha\beta i}^\ast \frac{\mu_i^{}}{M_{\Delta_i}^{}} \left[ g_{\alpha j}^{}g_{\beta j}^{}\sum_{j} I_{\Delta_i}^{N_j} +\sum_{k\neq i}^{} f_{\alpha\beta k}^{} \frac{\mu_k^{}}{M_{\Delta_i}^{}}I_{\Delta_i}^{\Delta_k}
+\sum_{l}^{}h_{\alpha l} h_{\beta l}^{} I_{N_i}^{T_l}  \right]\right\}}{\sum_{\alpha\beta}^{}f_{\alpha\beta i}^\ast f_{\alpha\beta i}^{} +\frac{\mu_i^2}{M_{\Delta_i}^2}}\,,\\
[2mm]
\label{cpa-iii}
\varepsilon_{T_i}^{}&=&\frac{1}{8\pi}\frac{\sum_{\alpha \beta}^{} \textrm{Im}\left\{h_{\alpha i}^\ast  h_{\beta i}^\ast \left[ g_{\alpha j}^{}g_{\beta j}^{}\sum_{j} I_{N_i}^{N_j} +\sum_{k}^{} f_{\alpha\beta k}^{} \frac{\mu_k^{}}{2M_{T_i}^{}}  I_{N_i}^{\Delta_k} +\sum_{l\neq i }^{}h_{\alpha l} h_{\beta l}^{} I_{N_i}^{T_l}  \right] \right\}}{\sum_{\alpha}^{}h_{\alpha i}^\ast h_{\alpha i}^{} }\,,
\end{eqnarray}
\end{widetext}
where the functions $I_{F_i}^{F_j}$, $I_{F_i}^{\Delta_j}$, $I_{\Delta_i}^{\Delta_j}$ and $I_{\Delta_i}^{F_j}~(F_i^{}=N_i^{}/T_i^{})$ are calculated by

\begin{eqnarray}
I_{F_i}^{F_j}\!\!&\equiv& \!\!I_{F_i}^{F_j}\left[\frac{M_{F_i}^2}{M_{F_j}^2}\right]~~~\textrm{with}\nonumber\\
[1mm]
&&\!\!I_{F_i}^{F_j}[x]\!=\!\frac{\sqrt{x}}{1-x}+\frac{1}{\sqrt{x}}\!\left[-1+\!\left(1+\frac{1}{x}\right)\!\ln\!\left(1+x\right)\!\right]\,,\nonumber\\
[1mm]
I_{F_i}^{\Delta_j}\!\!&\equiv& \!\!I_{F_i}^{\Delta_j}\left[\frac{M_{F_i}^2}{M_{\Delta_j}^2}\right] ~\textrm{with}\nonumber\\
[1mm]
&&\!\! I_{F_i}^{\Delta_j}[x]=\!3\left[1-\frac{1}{x}\ln\left(1+x\right)\right]\,,\nonumber\\
[2mm]
I_{\Delta_i}^{\Delta_j}\!\!&\equiv&\!\! I_{\Delta_i}^{\Delta_j}\left[\frac{M_{\Delta_i}^2}{M_{\Delta_j}^2}\right] ~\textrm{with}~ I_{\Delta_i}^{\Delta_j}[x]=\!\frac{2x}{1-x}\,,\nonumber\\
[2mm]
I_{\Delta_i}^{F_j}\!\!&\equiv&\!\! I_{\Delta_i}^{F_j}\left[\frac{M_{\Delta_i}^2}{M_{F_j}^2}\right] ~~\textrm{with}~  I_{\Delta_i}^{F_j}[x]=\!\frac{2}{\sqrt{x}}\ln\left(1+x\right)\,.
\end{eqnarray}

We find the CP violation $\varepsilon_{N_i/\Delta_i/T_i}^{}$ can have an exact or approximate dependence on the imaginary part of the neutrino mass matrix in some cases. Actually we read
\begin{itemize}
\item in the type-I seesaw with two fermion singlets,
\begin{subequations}
 \begin{eqnarray}
\varepsilon_{N_1}^{}\!\!&=&\!\! \frac{1}{8\pi}\frac{\sum_{\alpha \beta }g_{\alpha 1}^{} g_{\beta 1}^{}\textrm{Im}\left[\left(m_\nu^{}\right)_{\alpha\beta}^{}\right]M_{N_2}^{} I^{N_2}_{N_1} }{v^2_{}\sum_{\alpha}^{}g_{\alpha 1}^2}\,,\\
[2mm]
\varepsilon_{N_2}^{}\!\!&=&\!\!-\frac{1}{8\pi}\frac{\sum_{\alpha \beta }g_{\alpha 1}^{} g_{\beta 1}^{}\textrm{Im}\left[\left(m_\nu^{}\right)_{\alpha\beta}^{}\right]M_{N_1}^{} I^{N_1}_{N_2}}{v^2_{}\sum_{\alpha}^{}g_{\alpha 2}^{\ast} g_{\alpha 2}^{}}\,,
\end{eqnarray}
\end{subequations}
\item in the type-III seesaw with two fermion triplets,
\begin{subequations}
 \begin{eqnarray}
\varepsilon_{T_1}^{}\!\!&=&\!\! \frac{1}{8\pi}\frac{\sum_{\alpha \beta }h_{\alpha 1}^{} h_{\beta 1}^{}\textrm{Im}\left[\left(m_\nu^{}\right)_{\alpha\beta}^{}\right]}{\sum_{\alpha}^{}h_{\alpha 1}^2}\frac{M_{T_2}^{} I^{T_2}_{T_1}}{v^2_{}}\,,\\
[2mm]
\varepsilon_{T_2}^{}\!\!&=&\!\!-\frac{1}{8\pi}\frac{\sum_{\alpha \beta }h_{\alpha 1}^{} h_{\beta 1}^{}\textrm{Im}\left[\left(m_\nu^{}\right)_{\alpha\beta}^{}\right]}{\sum_{\alpha}^{}h_{\alpha 2}^{\ast}h_{\alpha 2}^{}}\frac{M_{T_1}^{} I^{T_1}_{T_2}}{v^2_{}}\,,
\end{eqnarray}
\end{subequations}
\item in the type-I+III seesaw with one fermion singlet and one fermion triplet,
\begin{subequations}
\begin{eqnarray}
\varepsilon_{N_1}^{}\!\!&=&\!\!\frac{1}{8\pi}\frac{\sum_{\alpha \beta }g_{\alpha 1}^{} g_{\beta 1}^{}\textrm{Im}\left[\left(m_\nu^{}\right)_{\alpha\beta}^{}\right] M_{T_1}^{}I_{N_1}^{T_1}}{v^2_{}\sum_{\alpha}^{}g_{\alpha 1}^2}\,,\\
[2mm]
\varepsilon_{T_1}^{}\!\!&=&\!\!-\frac{1}{8\pi}\frac{\sum_{\alpha \beta }g_{\alpha 1}^{} g_{\beta 1}^{}\textrm{Im}\left[\left(m_\nu^{}\right)_{\alpha\beta}^{}\right] M_{N_1}^{}I_{T_1}^{N_1}}{v^2_{}\sum_{\alpha}^{}h_{\alpha 1}^{\ast} h_{\alpha 1}^{}} \,,
\end{eqnarray}
\end{subequations}
\item in the type-I+II seesaw with one fermion singlet and one Higgs triplet,
\begin{subequations}
\begin{eqnarray}
\varepsilon_{N_1}^{}\!\!&=&\!\!\frac{1}{8\pi}\frac{\sum_{\alpha \beta }g_{\alpha 1}^{} g_{\beta 1}^{}\textrm{Im}\left[\left(m_\nu^{}\right)_{\alpha\beta}^{}\right] M_{\Delta_1}^{}I_{N_1}^{\Delta_1}}{v^2_{}\sum_{\alpha}^{}g_{\alpha 1}^2}\,,\\
[2mm]
\varepsilon_{\Delta_1}^{}\!\!&=&\!\!-\frac{1}{8\pi}\frac{\sum_{\alpha \beta }g_{\alpha 1}^{} g_{\beta 1}^{}\textrm{Im}\left[\left(m_\nu^{}\right)_{\alpha\beta}^{}\right] M_{N_1}^{}I^{N_1}_{\Delta_1}}
{v^2_{}\left(\sum_{\alpha\beta}^{}f_{\alpha\beta 1}^\ast f_{\alpha\beta 1} + \frac{\mu_1^2}{M_{\Delta_1}^2}\right)}\,,
\end{eqnarray}
\end{subequations}
\item in the type-III+II seesaw with one fermion triplet and one Higgs triplet,
\begin{subequations}
\begin{eqnarray}
\varepsilon_{T_1}^{}\!\!&=&\!\!\frac{1}{8\pi}\frac{\sum_{\alpha \beta }h_{\alpha 1}^{} h_{\beta 1}^{}\textrm{Im}\left[\left(m_\nu^{}\right)_{\alpha\beta}^{}\right] M_{\Delta_1}^{}I_{T_1}^{\Delta_1}}{v^2_{}\sum_{\alpha}^{}h_{\alpha 1}^2}\,,\\
[2mm]
\varepsilon_{\Delta_1}^{}\!\!&=&\!\!-\frac{1}{8\pi}\frac{\sum_{\alpha \beta }h_{\alpha 1}^{} h_{\beta 1}^{}\textrm{Im}\left[\left(m_\nu^{}\right)_{\alpha\beta}^{}\right] M_{T_1}^{}I^{T_1}_{\Delta_1}}
{v^2_{}\left(\sum_{\alpha\beta}^{}f_{\alpha\beta 1}^\ast f_{\alpha\beta 1} + \frac{\mu_1^2}{M_{\Delta_1}^2}\right)}\,.
\end{eqnarray}
\end{subequations}
\end {itemize}
When the above special seesaw models are extended by more fermion singlet(s)/triplet(s) and Higgs triplet(s), we can expect a leptogensis by the decays of the lightest fermion singlet/triplet. In this case, we can denote the lightest fermion singlet/triplet by $N_{1}^{}/T_{1}^{}$ and then consider the assignment (\ref{real}) in Eqs. (\ref{cpa-i}) and (\ref{cpa-iii}). The CP violation then can be simplified as
\begin{eqnarray}
\label{simple-i/iii}
\varepsilon_{N_1}^{}&=&\frac{3}{16\pi}\frac{\sum_{\alpha\beta}^{}\left\{g_{\alpha 1}^{} g_{\beta 1}^{} \textrm{Im}\left[\left(m_\nu^{}\right)_{\alpha\beta}^{}\right] \right\}M_{N_1}^{}}{ v^2_{}\sum_{\alpha}^{}g_{\alpha 1}^2}\quad \textrm{or}\nonumber\\
[2mm]
\varepsilon_{T_1}^{}&=&\frac{3}{16\pi}\frac{\sum_{\alpha\beta}^{}\left\{h_{\alpha 1}^{} h_{\beta 1}^{} \textrm{Im}\left[\left(m_\nu^{}\right)_{\alpha\beta}^{}\right] \right\}M_{T_1}^{}}{v^2_{}\sum_{\alpha}^{}h_{\alpha 1}^2 }\,,
\end{eqnarray}
which is easy to give us an upper bound \cite{di2002,bdp2003}.

Ones may be interested in the so-called Davidson-Ibarra parametrization \cite{di2001}, under which the Yukawa couplings $g/h$ in the pure type-I/III seesaw are determined by
\begin{eqnarray}
\label{par}
g_{\alpha i}^{}&=&i \sum_{j}^{}U_{\alpha j}^{} \sqrt{m_j^{}} O_{ji}^{} \sqrt{M_{N_i}^{}} /v \quad\textrm{or} \nonumber\\
[2mm]
 h_{\alpha i}^{}&=&i \sum_{j}^{}U_{\alpha j}^{} \sqrt{m_j^{}} O_{ji}^{} \sqrt{M_{T_i}^{}} /v\,,
\end{eqnarray}
with $O$ being an arbitrary complex orthogonal matrix. Ones hence conclude that in the presence of the complex orthogonal matrix $O$, the Yukawa couplings $g/h$ can be complex even if the PMNS matrix $U$ does not contain any CP phases. The CP asymmetry (\ref{simple-i/iii}) then can be irrelevant to the PMNS matrix, i.e.
\begin{eqnarray}
\label{di-i/iii}
\varepsilon_{N_1}^{}&=&-\frac{3}{16\pi}\frac{\textrm{Im}\left\{\sum_{j\neq 1}^{}\left(\sum_{k}^{}O_{k1}^\ast O_{kj}^{}m_k^{}\right)^2_{} \right\}M_{N_1}^{}}{ v^2_{}\sum_{k}^{}\left|O_{k1}^{}\right|^2_{}m_k^{}}\quad \textrm{or}\nonumber\\
[2mm]
\varepsilon_{T_1}^{}&=&-\frac{3}{16\pi}\frac{\textrm{Im}\left\{\sum_{j\neq 1}^{}\left(\sum_{k}^{}O_{k1}^\ast O_{kj}^{}m_k^{}\right)^2_{} \right\}M_{T_1}^{}}{ v^2_{}\sum_{k}^{}\left|O_{k1}^{}\right|^2_{}m_k^{}}\,.
\end{eqnarray}
Under our assignment (\ref{real}), the CP asymmetry (\ref{simple-i/iii}) depends on a complex diagonal matrix $X$ besides the PMNS matrix $U$, see Eqs. (\ref{real2}) and (\ref{numass}). Clearly, our $X$ matrix is very simple, compared with the $O$ matrix.

\underline{Conclusion:} In this work we have revealed the origin of the CP violation for the leptogenesis in the most popular seesaw models. Specifically, we find that in a pure type-I/III seesaw with two fermion singlets/triplets, a combined type-I+III seesaw with one fermion singlet and one fermion triplet, or a combined type-I/III+II seesaw with one fermion singlet/triplet and one Higgs triplet, the Yukawa couplings involving one of the two fermion 
singlets/triplets or the unique Higgs triplet are always allowed to absorb all of the physical CP phases in the lepton sector. In this basis, the CP violation required by the leptogenesis should exactly come from the imaginary part of the neutrino matrix. We also consider a generalization in the case that these seesaw scenarios are extended by more fermion singlets/triplets and Higgs triplets while the leptogenesis is realized by the decays of the lightest fermion singlet/triplet. This generalization is a very good approximation and is reliable even in the radiative type-I/III and type-I+III seesaw \cite{ma2006,kms2006} where an inert Higgs doublet, rather than the SM Higgs doublet, is responsible for the Yukawa couplings of the fermion singlet(s)/triplet(s) to the SM lepton doublets.

\textbf{Acknowledgement}: This work was supported by the Recruitment Program for Young Professionals under Grant No. 15Z127060004, the Shanghai Jiao Tong University under Grant No. WF220407201, the Shanghai Laboratory for Particle Physics and Cosmology under Grant No. 11DZ2260700 and the Key Laboratory for Particle Physics, Astrophysics and Cosmology, Ministry of Education.

\end{document}